\documentclass[runningheads]{llncs}

\usepackage[normalem]{ulem}
\usepackage[table,xcdraw]{xcolor}
\usepackage{color,soul}
\usepackage{dirtytalk}
\usepackage{ragged2e}
\usepackage{rotating}

\usepackage{graphicx}
\usepackage{amsmath}
\usepackage{amssymb}
\usepackage{booktabs}
\usepackage{textcomp}
\usepackage{xcolor}
\usepackage{graphics}
\usepackage{booktabs}
\usepackage{tabularx}
\usepackage{float}

\usepackage{times}
\usepackage{epsfig}
\usepackage{comment}
\usepackage{multirow}
\usepackage{pifont}
\usepackage{caption}
\usepackage{subcaption}
\usepackage{url}
\usepackage{soul}

\usepackage[pagebackref,breaklinks,colorlinks,bookmarks=false]{hyperref}

\usepackage[capitalize]{cleveref}
\crefname{section}{Sec.}{Secs.}
\Crefname{section}{Section}{Sections}
\Crefname{table}{Table}{Tables}
\crefname{table}{Tab.}{Tabs.}

\begin{document}
\title{Assessing the performance of deep learning-based models for prostate cancer segmentation using uncertainty scores}

\author{Pablo Cesar Quihui-Rubio\inst{1}, Daniel Flores-Araiza \inst{1}, Gilberto Ochoa-Ruiz\inst{1}, \\ Miguel Gonzalez-Mendoza\inst{1},   and Christian Mata\inst{2}\inst{,3} }
\authorrunning{P. Quihui-Rubio et al.}

\institute{Tecnologico de Monterrey, School of Engineering and Sciences, Mexico. 
\and Universitat Politècnica de Catalunya, 08019 Barcelona. Catalonia, Spain.
\and Pediatric Computational Imaging Research Group, Hospital Sant Joan de Déu, Esplugues de Llobregat, 08950, Catalonia, Spain \\}

\maketitle              

\begin{abstract}

This study focuses on comparing deep learning methods for the segmentation and quantification of uncertainty in prostate segmentation from MRI images. The aim is to improve the workflow of prostate cancer detection and diagnosis. Seven different U-Net-based architectures, augmented with Monte-Carlo dropout, are evaluated for automatic segmentation of the central zone, peripheral zone, transition zone, and tumor, with uncertainty estimation. The top-performing model in this study is the Attention R2U-Net, achieving a mean Intersection over Union (IoU) of $76.3\% \pm 0.003$ and Dice Similarity Coefficient (DSC) of $85\% \pm 0.003$ for segmenting all zones. Additionally, Attention R2U-Net exhibits the lowest uncertainty values, particularly in the boundaries of the transition zone and tumor, when compared to the other models.
\end{abstract}

\section{Introduction}

Prostate cancer (PCa) is the most common solid non-cutaneous cancer in men and is among the most common causes of cancer-related deaths in 13 regions of the world \cite{ferlay2021}. According to a recent overview, in 2020 prostate cancer was the most frequently diagnosed cancer in males in 12 regions of the world, which translates to around 1.41 million new cases \cite{ferlay2021}. However, when detected in early stages, the survival rate for regional PCa is almost 100\%. In contrast, the survival rate when the cancer is spread to other parts of the body is of only 30\% \cite{astrazeneca_2020}.

Magnetic Resonance Imaging (MRI) is the most widely available non-invasive and sensitive tool for detection, localization and staging of PCa, due to its high resolution, excellent spontaneous contrast of soft tissues, and the possibility of multi-planar and multi-parametric scanning \cite{Chen_2008}. MRI can be also be used for PCa detection through the segmentation of Regions of Interest (ROI).
The use of image segmentation for PCa can help determine the localization and the volume of the cancerous tissue \cite{Haralick_1985}. Although prostate image segmentation is a relatively old problem and some novel methods have been proposed, radiologists still perform a manual segmentation of the prostate gland and regions of interest (central zone, peripheral zone, and transition zone) \cite{aldoj2020}. This manual process is time-consuming, and is sensitive to the specialist experience, resulting in a significant intra- and inter-specialist variability. Therefore, automating the process of segmentation of prostate and gland regions of interest, may help save time for practitioner radiologists and additionally can be used as a training tool for others. 
One of the most popular architectures is the U-Net \cite{Unet_ronneberger_2015} model, which has been the inspiration behind many recent works in the literature, such as Swin U-Net \cite{swinunet}, or R2U-Net \cite{r2unet}. While these models have yielded positive outcomes, inconsistencies in performance have been observed in U-Net-based segmentation due to the prostate's anatomical structure. The boundaries between zones can distort semantic features, leading to unreliable results. Furthermore, automatic segmentation typically produces deterministic segmentation outcomes \cite{pone}, and there is insufficient information available about the model's confidence level \cite{yonkai2020uncertainty}.
Despite their successes in many medical image analysis applications, DL algorithms are usually not translated into real-world clinical scenarios because these do not provide information about the uncertainty associated with their prediction. This is problematic in the challenging context of pathological structures segmentation (e.g, tumors) as even the top-performing methods are prone to errors, and due to the lack of uncertainty information, it results impossible tell apart different sorts of erroneous predictions.

Therefore, the overall segmentation workflow can be improved by providing the uncertainties of the model that could allow end-users (e.g, clinicians) to review and refine cases with high uncertainty.

In this work, we carry out a thorough assessment of automatic prostate zone segmentation models using U-Net, Attention U-Net, Dense U-Net, Attention Dense U-Net, R2U-Net, Attention R2U-Net, and Swin U-Net architectures. Additional to the segmentation task, we include the pixel-wise estimation of the uncertainty, which can be done by obtaining a probability distribution of the weights of the model.
The zones evaluated in this work are the central zone (CZ), the peripheral zone (PZ), transition zone (TZ), and, in the case of a disease, the tumor zone (TUM), unlike previous works which only evaluate CZ and PZ \cite{yonkai2020uncertainty}.

%
This paper has five sections including this introduction. Section \ref{sec:related_work} provides a review about what has been done in previous works related to prostate segmentation and uncertainty quantification. Section \ref{sec:methods} the dataset used is described, followed by a description of the uncertainty quantification procedure in this segmentation task. In section \ref{sec:results} the results of the experiments are discussed in detail. Finally the conclusion of this work is presented in Section \ref{sec:conclusion}. 

\section{Related Work}
\label{sec:related_work}
\subsection{Deep Learning Segmentation}


For segmentation, one of the best known models in the literature is the U-Net architecture \cite{Unet_ronneberger_2015}, which is the base for many other novel models. The work from Zhu et \textit{al.} \cite{Zhu_deeplycnn_2017} proposes a U-Net based network to segment the whole prostate gland, obtaining encouraging results (DSC of 0.885). Moreover, this architecture has served as the inspiration for some variants that enhance the performance of the original model. One example is the work from Clark et \textit{al.} \cite{clark_2017} that presents a model that combines concepts from the U-Net and the inception architectures. Another example is the work presented by Oktay et \textit{al.} \cite{Att-unet-2018}, which proposes the addition of attention gates inside the original U-Net model with the intention of focusing on specific target structures. The addition of attention has served as base for other architectures such as Attention Dense U-Net \cite{Att-denseunet-2019}, Attention R2U-Net \cite{r2unet}, among others. Also, the introduction of Transformers in U-Net architectures is a novel approach for segmentation task that had demonstrated a good performance in biomedical images, such as Swin U-Net \cite{swinunet}. Despite this, during the course of this study, no other research was found that segmented the four zones discussed in this paper. Therefore, the number of studies that consider a third zone (TZ) is still limited, this is more likely because the most common datasets used are PROMISE-12 and the one from the PROSTATEx challenge, with only CZ and PZ. In addition to that, providing a value that quantifies the uncertainty of the predictions can improve the overall workflow since it could easily allow refining uncertain cases by human experts.


\subsection{Uncertainty Quantification}

The work from Theckel et \textit{al.} \cite{tinu2021analyzing} introduces a U-Net architecture with spatial dropout to measure the uncertainty related to the segmentation of macular degeneration, utilizing different sizes of input data. The work from Suman et \textit{al.} \cite{suman2018retina} applied the uncertainty quantification problem to retinal imaging using a ResNet-based model, modified with standard random dropout layers before every convolutional block. The work from Liu et \textit{al.} \cite{yonkai2020uncertainty} proposes an automatic segmentation of the prostate zones and introduces a pixel-wise uncertainty estimator using a ResNet50 backbone with attention and dropout layers.

\section{Materials and Methods}
\label{sec:methods}

\subsection{Dataset}
\label{sec:dataset}


The dataset used in the present work was provided by \textit{Universidad Politécnica de Cataluña} (UPC) in Barcelona, and Centre Hospitalaire de Dijon in France. The dataset consists of three-dimensional T2-weighted fast spin-echo (TR/TE/ETL: 3600 ms/ 143 ms/109, slice thickness:$1.25$ mm) images acquired with sub-millimeter pixel resolution in an oblique axial plane. The number of patients in the dataset are 19, with a total of 205 images with their corresponding annotation masks (of prostate zones) used as ground truth which were validated by experts using a dedicated tool \cite{Mata_2022}. 


The full dataset of 205 images, contains four different combination of zones, being: (CZ+PZ), (CZ+PZ+TZ), (CZ+PZ+Tumor), and (CZ+PZ+TZ+Tumor) with 73, 68, 23, and 41 images, respectively. For the purpose of this work, the dataset was divided in 85\% for training and 15\% for testing. 

\subsection{Uncertainty Estimation in Prostate Segmentation}


Epistemic and aleatory uncertainties are the two major types of uncertainty that can be quantified. Epistemic uncertainty captures the uncertainty related to the models parameters caused by the lack of data, and, aleatory uncertainty captures the noise inherent in the input data \cite{yonkai2020uncertainty}. The sum of both uncertainties forms the predictive uncertainty.


In this work, the uncertainty of seven different U-net-based models was measured in the test set. To approximate the inference of a model, Monte Carlo (MC) dropout of a hidden layer was performed. MC Dropout is a technique used in neural networks to incorporate uncertainty. It treats a network with dropped-out neurons as Monte Carlo samples from all possible combinations, approximating a Gaussian process \cite{yonkai2020uncertainty,gal2016dropout}. The minimization of cross-entropy loss is similar to minimizing the divergence of the predicted distribution \cite{suman2018retina}. Using MC Dropout, pixel-wise epistemic uncertainty can be computed as a variational Bayesian inference problem \cite{suman2018retina}. During predictions or testing, dropout is also necessary. The main focus of this study is to investigate the predictive uncertainty of prostate segmentation, which can be quantified using the entropy of the predictive distribution \cite{yonkai2020uncertainty}.

\subsection{Proposed Work}

This work uses the original U-Net model and six U-Net extensions from the literature: Attention U-Net \cite{Att-unet-2018}, Dense U-Net \cite{dense_unet_wu_2021}, Attention Dense U-Net \cite{Att-denseunet-2019}, R2U-Net \cite{r2unet}, Attention R2U-Net, and Swin U-Net \cite{swinunet}. These architectures had demonstrated great performance segmenting biomedical images, even some of them with public prostate's datasets including CZ and PZ. However, unlike in other works, we proposed to compare the performance segmenting the three main zones of the prostate (CZ, PZ, and TZ) and a tumor tissue if it is present, using the dataset described in Section \ref{sec:dataset}. 



Before the final training, an hyperparameter tunning proccess using a stratified 5-Fold validation with the training set was carried out using the base U-Net model in order to obtain the optimal combination of data augmentation, learning rate and an approximation of epochs for training. The results demonstrated that including data augmentation in the training did not increase significantly the performance of the models. Therefore we decided to use the original dataset without data augmentation due to computational resources and time processing. The previously mentioned models were trained for 145 epochs, using Adam optimizer with a learning rate of $1e-4$ and Categorical Cross-Entropy (CCE) loss function. The performance was evaluated using Dice Score (DSC) and Intersection over Union (IoU) as the main metrics.

\section{Results and Discussion}
\label{sec:results}



\subsection{Quantitative Results}
\vspace{-0.15cm}

Table \ref{tab:results_seg} shows a summary of evaluation results of the seven studied architectures, in terms of two metrics (DSC and IoU) and loss value. In order to obtain these results, the evaluation of each model was performed $T = 50$ times, and due to the incorporation of MC Dropouts the results were different each time. Therefore, the average of all evaluations and prostate zones is reported with their corresponding standard deviation.

\begin{table}[htp]
\vspace{-0.15cm}
\centering
\caption{Comparison of model performance in segmentation metrics and loss value. The metrics are denoted by upward ($\uparrow$) or downward ($\downarrow$) arrows, indicating the desired direction of values. Bold values highlighted in green represent the best score achieved among all models.}
\label{tab:results_seg}
\begin{tabular}{@{}l|c|c|c@{}}
\toprule
\multicolumn{1}{c|}{Model} & IoU $\uparrow$    & DSC $\uparrow$    & Loss $\downarrow$\\ \midrule
U-Net                      & 0.676 $\pm$ 0.021 & 0.770 $\pm$ 0.021 & 0.0139 $\pm$ 0.0007 \\
Attention U-Net            & 0.688 $\pm$ 0.011 & 0.781 $\pm$ 0.010 & 0.0132 $\pm$ 0.0003 \\
Swin U-Net                 & 0.725 $\pm$ 0.014 & 0.816 $\pm$ 0.014 & 0.0134 $\pm$ 0.0002 \\
Dense U-Net                & 0.754 $\pm$ 0.004 & 0.846 $\pm$ 0.004 & 0.0146 $\pm$ 0.0003 \\
Attention Dense U-Net      & 0.760 $\pm$ 0.006 & 0.847 $\pm$ 0.005 & 0.0154 $\pm$ 0.0004 \\
R2U-Net                    & \cellcolor[HTML]{D8FFD8}\textbf{0.764 $\pm$ 0.002} & \cellcolor[HTML]{D8FFD8}\textbf{0.850 $\pm$ 0.002} & 0.0119 $\pm$ 0.0001 \\
Attention R2U-Net          & 0.763 $\pm$ 0.003 & \cellcolor[HTML]{D8FFD8}\textbf{0.850 $\pm$ 0.003} & \cellcolor[HTML]{D8FFD8}\textbf{0.0113 $\pm$ 0.0001} \\
\bottomrule
\end{tabular}
\end{table}

Based on the metrics values, it can be seen that U-Net was the model with worst performance. The use of attention to focus on the ROI helped to slightly outperform the performance in segmentation tasks compared to the original U-Net by around $1-2\%$ for IoU and DSC.

Moving to Swin U-Net, a novel architecture from the state-of-the-art that uses Swin Transformers \cite{swinunet,swintransformer} achieved to increase the IoU and DSC values by more than $7\%$, and lower loss value compared to U-Net. 

In the case of Dense U-Net, the performance of the model exceeds the previous three architectures, with IoU and DSC scores $11\%$ and $10\%$ better than the base U-Net, respectively, with a loss value of $0.0146$. As a plus, this model did not need more computational resources or time during its training compared to base U-Net. The next model consisted on the incorporation of attention modules to Dense U-Net, which again outperformed all the previous models in the segmentation metrics by $12\%$ of IoU, and $10\%$ of DSC compared to U-Net. However, it achieved the higher loss value among all of $0.0154$.

The last two architectures R2U-Net and Attention R2U-Net achieved very similar results, but outperformed all the other models with values of $76.4\%$ and $85\%$ for IoU and DSC, respectively, and the lowest loss value of $0.0113$ for the Attention R2U-Net. 

\begin{figure}[htp]
\centering
\includegraphics[width=0.9\linewidth]{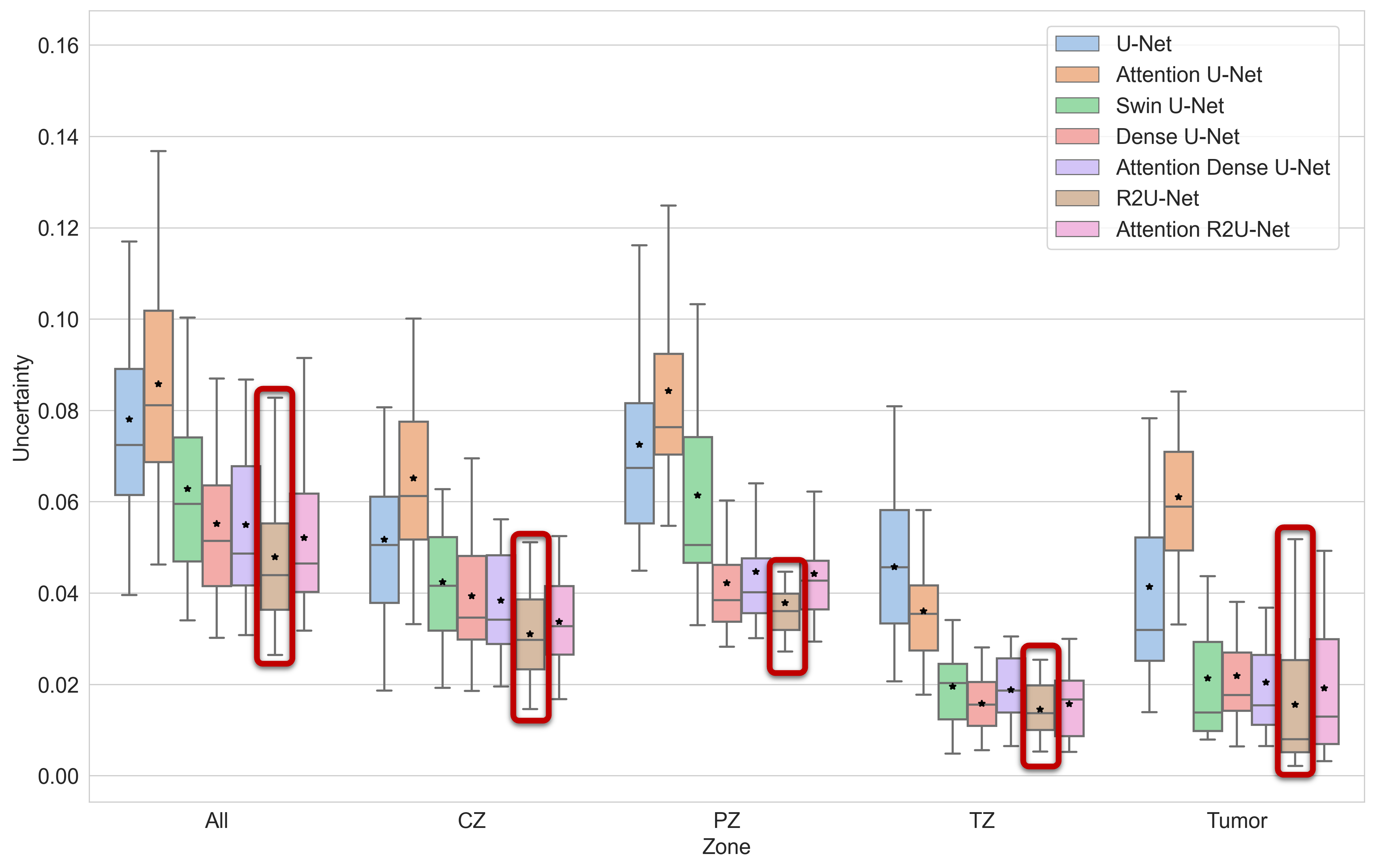}
\vspace{-0.35cm}
\caption{Comparison of Uncertainty per each class between DL Architectures. The mean uncertainty could be identify with a black star inside each box, and the line represents the median uncertainty value obtained, the best model is indicated with a red box for each zone.} 
\label{fig:boxplot}
\end{figure}

As mentioned before, an uncertainty comparison between the architectures was carried out per each prostate zone, as well as for the full image with its corresponding standard deviation as it is shown in Figure \ref{fig:boxplot}. The results shown in this figure can help us to determine, in relation with previous table, which model achieved to segment with more certain the prostate and its zones.

In Figure \ref{fig:boxplot} it is observed that overall, the model that had the lowest mean uncertainty segmenting all the images in the test set was R2U-Net with a mean value of $0.048 \pm 0.014$ after 50 predictions, validating the results obtained in the Table \ref{tab:results_seg}, being the most reliable and accurate model overall thanks to the use of recurrent and residual units to get more context information.

Furthermore, the Attention U-Net was the one with the highest uncertainty overall with a value of $0.086 \pm 0.023$, having poor results in comparison to the other models. U-Net and Swin U-Net obtained very similar results in most of the prostate zones, although in the case of the TZ and Tumor, Swin U-Net achieved lower uncertainty. 

Dense U-Net, Attention Dense U-Net and Attention R2U-Net succeeded in obtaining smaller uncertainty mean values than U-Net ($0.055 \pm 0.018$, $0.054 \pm 0.018$, and $0.052 \pm 0.014$, respectively). Although, TZ and Tumor are the zones less present in the dataset, and where it looks to be more complex to segment, models like R2U-Net and Attention R2U-Net managed to achieved a great segmentation performance and uncertainty values in average of those zones in the test set. It is important to notice that both results are correlated. These models managed to be adequately trained to perform the most accurate segmentation task among the others, which can give more confidence to radiologists when using a prostate segmentation tool based in this trained model.

\subsection{Qualitative Results}

\begin{figure*}[htp]
\centering
 \includegraphics[width=1\linewidth]{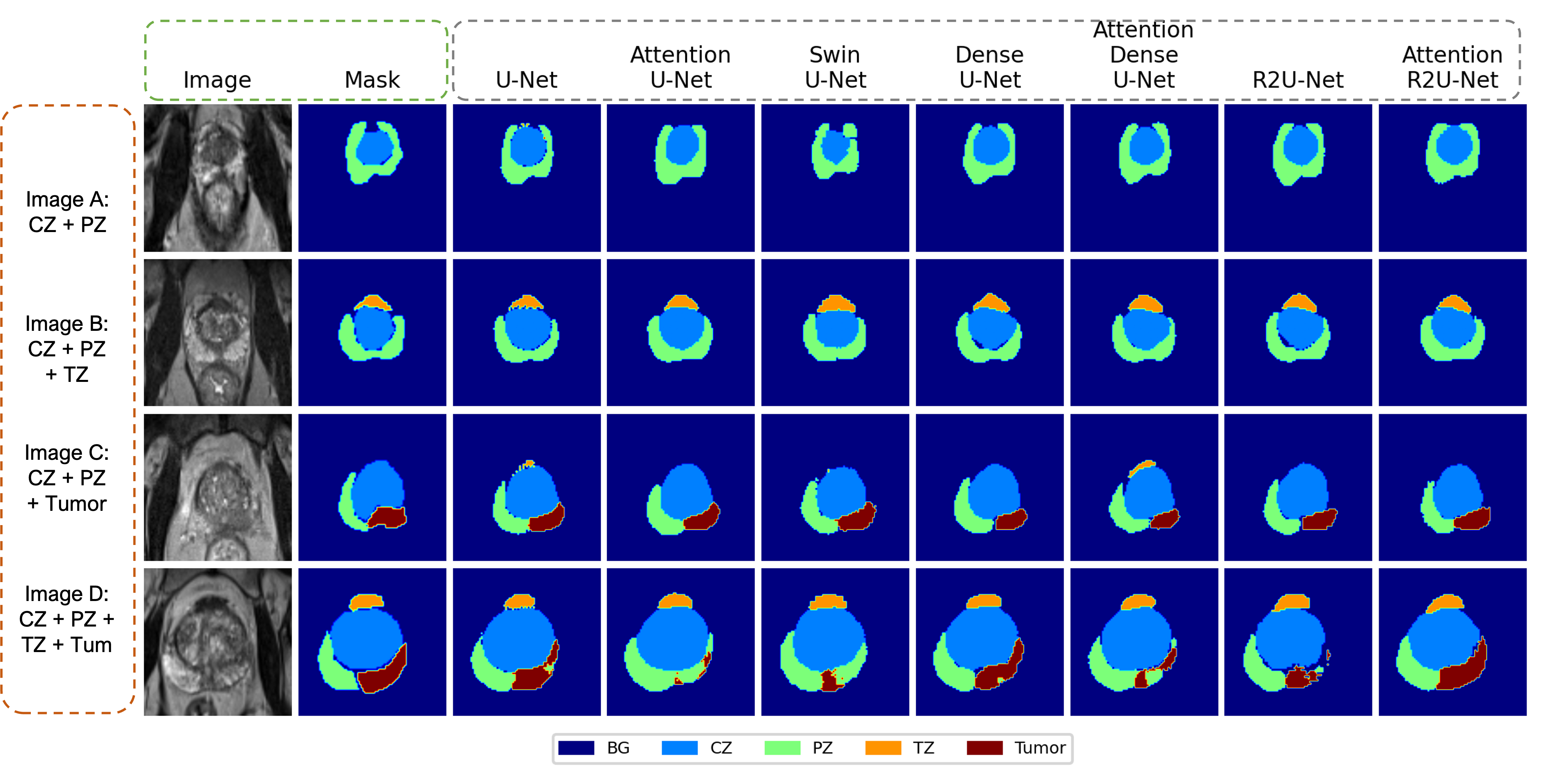}
\caption{Comparison of average segmentation after 50 predictions for each model in all the combinations of zones in the dataset.}
\label{fig:results1}
\end{figure*}

In Figure \ref{fig:results1}, a qualitative comparison is presented among the predictions of each model using four different example images from the dataset. The comparison involves all possible combinations of zones. The first two columns display the original T2-MRI image of the prostate and its corresponding ground truth mask. Subsequently, each column represents the average of probabilities obtained from 50 predictions for each model. It can be observed that the first two zone combinations (Image A and B in Figure \ref{fig:results1}) are relatively easier for most models, as they produce segmentation that closely resemble the ground truth. However, certain models such as U-Net and Swin U-Net appear to misclassify pixels as TZ even when they are not present in the ground truth. Nevertheless, based on the examples in the test set, the models have been trained effectively to achieve satisfactory segmentation performance on images containing CZ and PZ, and some including TZ. 

Regarding the other two combinations that include the tumor, they posed the most complex segmentation challenge with notable variation among models. In Image C of Figure \ref{fig:results1}, models like U-Net and Attention Dense U-Net incorrectly classified a TZ region that was not identified in the ground truth. Meanwhile, other models tended to excessively smooth the original segmentation, yielding a seemingly good but possibly inaccurate result. However, when visually compared to the ground truth, the best segmentation in this example was achieved by R2U-Net and Attention R2U-Net. 

\begin{figure*}[htp]
\centering
\includegraphics[width=1\linewidth]{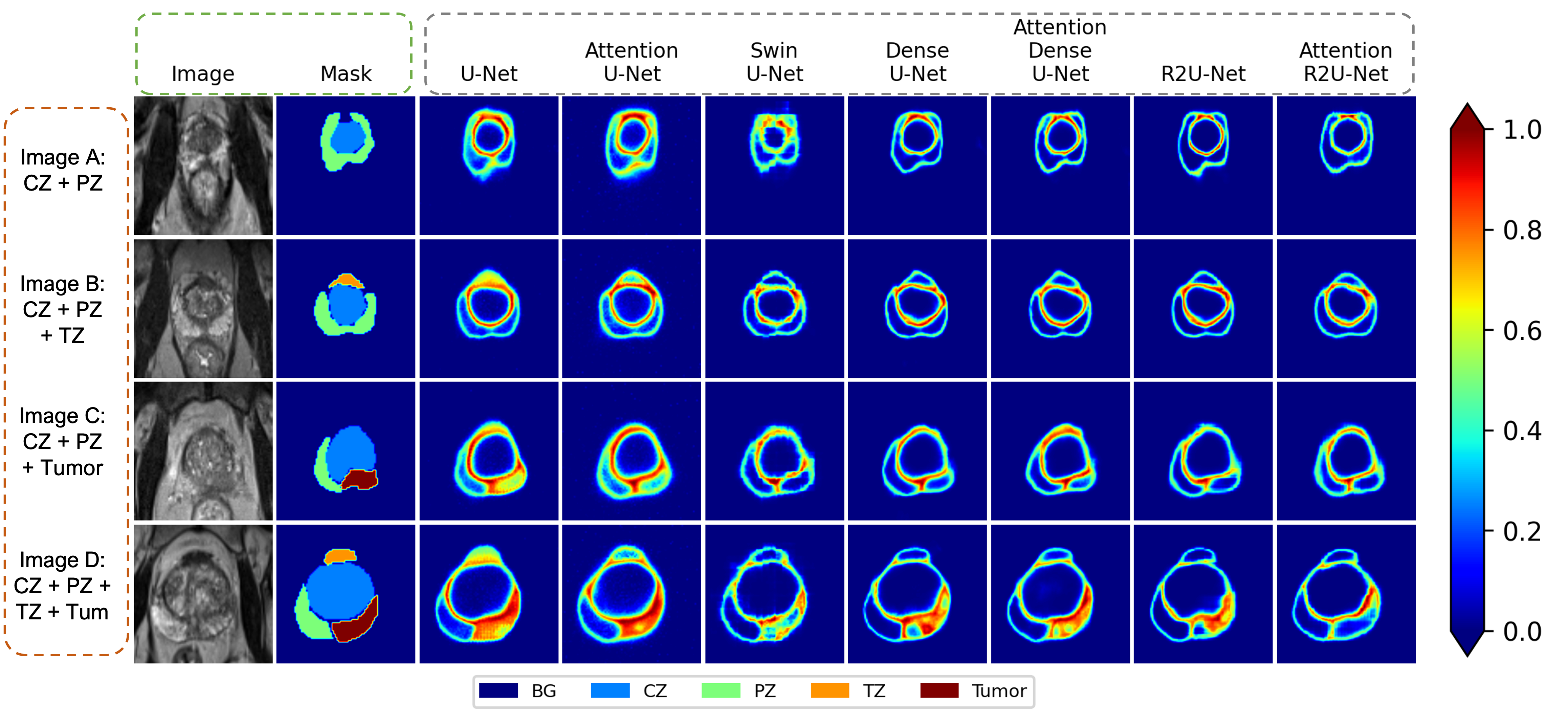}
\caption{Comparison of uncertainty maps after 50 predictions for each model with previous examples.}
\label{fig:results2}
\end{figure*}

For the last example, most models struggled to accurately segment the tumor. Surprisingly, U-Net and Dense U-Net produced acceptable results, but Attention R2U-Net demonstrated the best overall performance.

Figure \ref{fig:results2} illustrates the significance of uncertainty by displaying the same four examples as in the previous figure, along with corresponding uncertainty maps represented as heat maps for each trained model. The temperature of the image indicates the level of uncertainty, with higher temperatures indicating greater uncertainty in those pixels, while lower temperatures indicate higher certainty in the model's pixel segmentation.

The model with the highest uncertainty, particularly around the boundaries of TZ and tumor, is U-Net, followed by Attention U-Net. This observation is evident. Furthermore, as previously mentioned, the first two examples were easier for the models, resulting in relatively low uncertainty across most of them. When dealing with images containing tumors, the inclusion of dense blocks enhanced model certainty. However, the utilization of recurrent residual blocks and attention modules surpassed other models, achieving acceptable predictions in the test set with low uncertainty values, even in TZ and tumor tissues.

\section{Application}

\begin{figure*}[htp]
    \centering
    \includegraphics[width=0.7\linewidth]{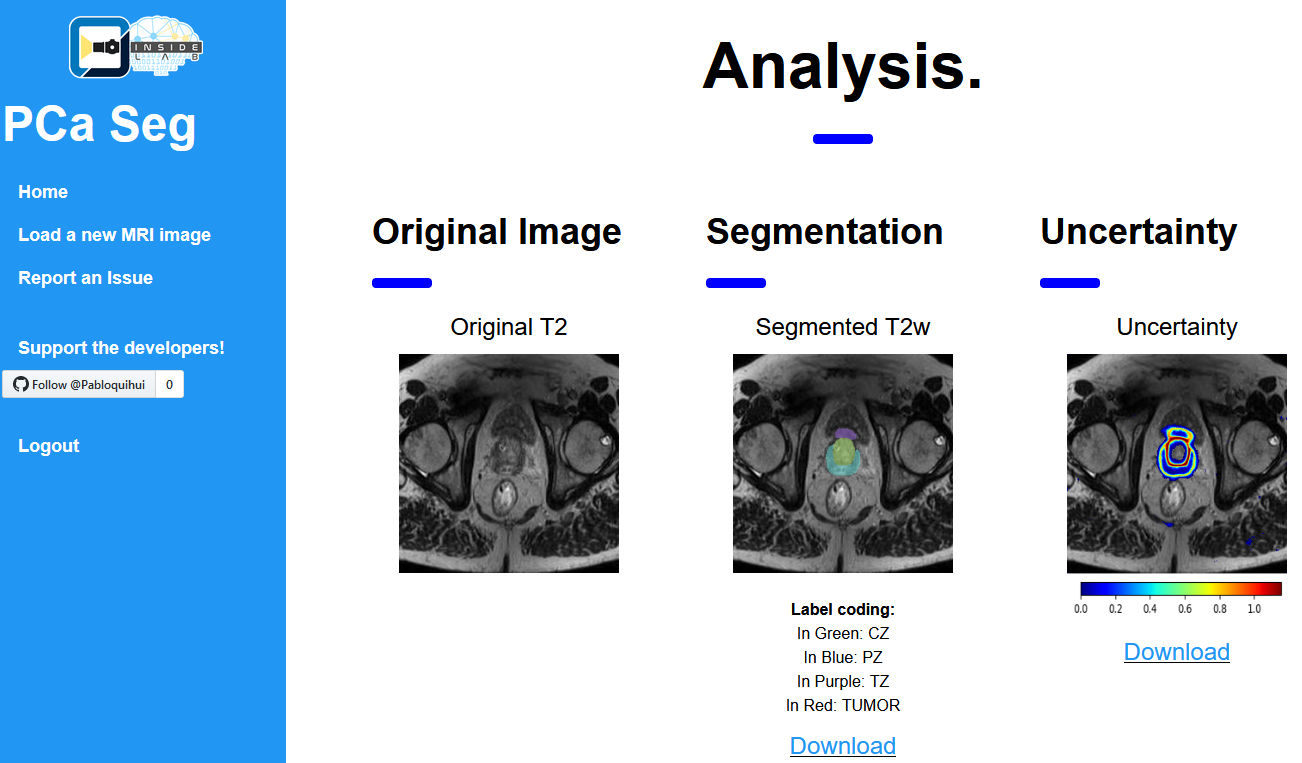}
    \caption{Example of the analysis page of the \textit{'ProstAI'} app using a prostate image from the Test set.} \label{fig:webapp}
\end{figure*}

In order to have a computer-aided tool which can be used for radiologists or clinicians, we proposed a Web App using Flask framework which we called \textit{'ProstAI'}, and it was designed to have easier access to predict images using the best trained model with MC dropouts: Attention R2U-Net. This app predicts the segmentation mask, as well as the uncertainty map, which is very helpful to indicate the experts which are the pixels where the model has higher uncertainty about their segmentation, an example is shown in Figure \ref{fig:webapp}.

This tool is proposed for experimental usage, further information about the app and an example of usage can be found in: \url{https://github.com/pabloquihui/ProstAI}.

\section{Conclusion}
\label{sec:conclusion}

This study makes a valuable contribution to prostate cancer segmentation by introducing the segmentation of transition and tumor zones, along with the quantification of uncertainty, which has received limited attention in existing literature. The utilization of a private dataset validated by multiple experts, including two radiologists and two oncologists, enhances the reliability and accuracy of the findings. A comparison of seven different deep learning models was conducted using segmentation metrics, uncertainty scores, and visual inspection. Among these models, Attention R2U-Net emerged as the top-performing approach in both analyses. The inclusion of recurrent residual blocks in U-Net (R2U-Net) notably enhanced the segmentation results by capturing additional contextual information. Furthermore, Attention R2U-Net demonstrated exceptional proficiency in segmenting all prostate zones, exhibiting superior performance in metrics and yielding lower average uncertainty estimated using the MC method. This highlights the positive impact of attention modules on improving segmentation and, more significantly, reducing uncertainty in predictions by focusing on the ROI.

\par Moreover, a web app has been developed with a focus on experimental use for radiologists. This app provides more accurate, consistent, and faster results and displays the uncertainty map for each predicted image. The uncertainty map provides a visual representation of the pixels in which the model is uncertain about the segmentation, giving radiologists a better idea of the areas that require further analysis. 

\section{Acknowledgments}

The authors wish to acknowledge the Mexican Council for Science and Technology (CONACYT) for the support in terms of postgraduate scholarships in this project, and the Data Science Hub at Tecnologico de Monterrey for their support on this project. 
This work has been supported by Azure Sponsorship credits granted by Microsoft's AI for Good Research Lab through the AI for Health program. The authors would also like to thank the financial support from Tecnologico de Monterrey through the “Challenge-Based Research Funding Program 2022”. Project ID 	\# E120 - EIC-GI06 - B-T3 - D.

\bibliographystyle{unsrt}
\bibliography{References.bib}
\end{document}